\documentclass{article-arxiv}

\journal{oe}

\usepackage{pifont}
\newcommand{\cmark}{\ding{51}}%
\newcommand{\xmark}{\ding{55}}%
\usepackage[utf8]{inputenc}
\usepackage{comment}
\usepackage{url}
\usepackage{newtxtext,newtxmath}

\begin{document}

\title{{\fontfamily{cmr}\selectfont High-speed object detection with a single-photon time-of-flight image sensor}}

\author{{\fontfamily{cmr}\selectfont Germán Mora-Martín,\authormark{1,*} Alex Turpin,\authormark{2,3} Alice Ruget,\authormark{4} Abderrahim Halimi,\authormark{4} Robert Henderson,\authormark{1} Jonathan Leach,\authormark{4} and Istvan Gyongy\authormark{1}}}

{\fontfamily{cmr}\selectfont
\address{\authormark{1}School of Engineering, Institute for Integrated Micro and Nano Systems, The University of Edinburgh, Edinburgh EH9 3FF, UK\\
\authormark{2}School of Computing Science, University of Glasgow, Glasgow G12 8QQ, UK\\
\authormark{3}ICFO-Institut de Ciencies Fotoniques, The Barcelona Institute of Science and Technology, Castelldefels, Barcelona, 08860, Spain\\
\authormark{4}School of Engineering and Physical Sciences, Heriot-Watt University, Edinburgh EH14 4AS, UK\\
\authormark{*}german.mora@ed.ac.uk}
}



\begin{abstract}
3D time-of-flight (ToF) imaging is used in a variety of applications such as augmented reality (AR), computer interfaces, robotics and autonomous systems. Single-photon avalanche diodes (SPADs) are one of the enabling technologies providing accurate depth data even over long ranges. By developing SPADs in array format with integrated processing combined with pulsed, flood-type illumination, high-speed 3D capture is possible. However, array sizes tend to be relatively small, limiting the lateral resolution of the resulting depth maps, and, consequently, the information that can be extracted from the image for applications such as object detection. In this paper, we demonstrate that these limitations can be overcome through the use of convolutional neural networks (CNNs) for high-performance object detection. We present outdoor results from a portable SPAD camera system that outputs 16-bin photon timing histograms with 64$\times$32 spatial resolution. The results, obtained with exposure times down to 2 ms (equivalent to 500 FPS) and in signal-to-background (SBR) ratios as low as 0.05, point to the advantages of providing the CNN with full histogram data rather than point clouds alone. Alternatively, a combination of point cloud and active intensity data may be used as input, for a similar level of performance. In either case, the GPU-accelerated processing time is less than 1 ms per frame, leading to an overall latency (image acquisition plus processing) in the millisecond range, making the results relevant for safety-critical computer vision applications which would benefit from faster than human reaction times.
\medskip
\end{abstract}

\section{Introduction}

 3D imaging is used in a wide range of applications, including computer interfaces, AR/VR \cite{AR/VR}, self-driving cars \cite{AutonomousD}, face recognition in smartphones \cite{Facerec}, robotics and ballistics \cite{Ballistics}. Common methods of retrieving depth data from a scene include stereoscopy or structured light, although these can be computationally expensive, and the accuracy reduces with distance. Furthermore, stereoscopy has difficulties with capturing scenes that are untextured \cite{Structured}. ToF is an alternative way to get depth information by using a modulated or pulsed light source (commonly a laser diode or a LED) typically emitting at near-infrared (NIR) wavelengths \cite{ToF} and measuring the time for the back-scattered light to return. ToF cameras usually reach speeds between 10-60 FPS and these can be divided in two groups: indirect ToF (iToF) and direct ToF (dToF). An iToF sensor measures the time-dependent intensity to retrieve the delay between transmitted and detected signal \cite{iToF}. Photo-demodulator pixels with multiple storage nodes are typically used, which are synchronised with the illumination source. iToF cameras are widely available commercially, and offer high lateral resolutions (e.g. 1024$\times$1024 \cite{Kinect}) and millimeter or sub-millimeter precision \cite{Lucid}. One of the main disadvantages is an inherent trade-off between range and accuracy, which typically limits the range to a few meters \cite{Range}. Scene-dependent, multi-path issues can also arise, leading to inaccurate depth data that can potentially confuse computer vision systems. Moreover, ambient illumination in outdoor use can also severely impact depth estimates \cite{Multipath}.

\medskip

Direct ToF sensors use a sub-nanosecond electronic stopwatch to measure the time for the light signal, typically a short laser pulse, to return. Highly sensitive avalanche photo-diodes (APD) or single-photon avalanche diodes (SPAD) are commonly used as detectors. dToF systems traditionally relied on single-point detectors, complemented by optical scanning to cover a given field of view (FoV), at the expense of a low acquisition rate. However, advances in SPAD technology have resulted in array format sensors with integrated timing electronics. SPAD-based dToF devices in line \cite{Line} and image sensor format \cite{Imagesensor1,Imagesensor2} using blade-scanned and flood illumination sources respectively, have been developed. These devices enable high-speed 3D imaging, especially when featuring on-chip processing such as histogramming \cite{SPAD}. Moreover, dToF sensors can maintain a high accuracy over hundreds of meters even under sunlight and do not suffer from multipath issues like iToF, making them the basis of LiDAR devices \cite{Lidar1,Lidar2}.

\medskip

Indeed, automotive LiDAR and autonomous driving have motivated a lot of research into both dToF sensors, and object detection based on depth data as considered here. Deep learning using convolutional neural networks (CNNs) reach higher levels of accuracy \cite{DL} with respect to older machine learning techniques like support vector machines (SVMs), k-nearest neighbours or gradient boosting trees \cite{SVM} in the context of object detection. High-resolution intensity/RGB images are the most common choice to perform object detection, as seen in popular neural networks such as ImageNet or YOLO \cite{YOLO}. LiDAR or point-based data has also been tested in RangeNet or PointNet networks \cite{DL-Lidar1,DL-Lidar2}, where raw point clouds are fed into the neural networks for 3D classification. More recent networks like 3DSSD \cite{3DSSD} have used the same idea to perform 3D object detection by processing efficiently point cloud data or PV-RCNN \cite{PVRCNN}, using a combination of voxels and point clouds.

\medskip

This paper targets low latency, CNN-based (U-net \cite{U-net}), object detection over short ranges (<10m) based on the output of a high-speed SPAD dToF image sensor. Instead of using point-cloud or voxel input (as typically done in literature), we investigate the impact of feeding raw histogram data to the neural network. Indeed, recent works have shown that the amount of information carried by temporal histograms in ToF measurements is significantly rich \cite{lindell2018single,Superres1,Sun:20}, even allowing for obtaining 3D image estimates from pure temporal data gathered with single-pixel time-resolving detectors \cite{Turpin:20,MonocularDepth}. Furthermore, we compare the performance of histogram-based processing with the result of using higher lateral resolution intensity (photon counting) data from the same sensor, as input to the CNN. The object detection is made more challenging by the presence of strong ambient light, as well as a level of occlusion in the dataset. The use of a U-net in this study as opposed to other state-of-the-art networks \cite{RGBD} that could potentially provide higher performance (e.g. YOLO) is based on the following: (1) U-net’s easily adaptable structure, which allows different data types (in particular the histogram data here) to be readily introduced (2) the ability to perform well even without large training datasets \cite{U-net} (3) moderate model sizes leading to fast processing speeds.

\section{SPAD sensor output}

We use a recently developed, reconfigurable SPAD image sensor, implemented in 3D-stacked technology \cite{SPAD}. The sensor contains 256$\times$256 SPADs which are grouped in 64$\times$64 macropixels, each containing 4$\times$4 SPADs. The readout is over 100 MHz output lines, giving a maximum frame rate of 760 FPS when the whole array is used and over 1000 FPS when half of it is used, as in the present study \cite{SPAD2}.

\medskip

The SPAD is able to operate in different modes. The modes used here are the intensity or photon counting mode (SPC), which runs at the SPAD resolution (256$\times$128) and time correlated single photon counting (TCSPC) mode, which operates at the macropixel resolution (64$\times$32). The latter mode generates 16-bin photon timing histograms per macropixel with 14-bit depth and a minimum temporal resolution of 500 ps per bin (the temporal bin width being programmable). If the laser pulse is spread over multiple bins and there are sufficient photon counts in the histogram, depth can be estimated with sub-bin precision. With the present sensor, sub-centimeter precision is achievable. Then, depth is estimated from each histogram with low computational cost by performing a centre-of-mass calculation following Eq. (\ref{Eq:1})

\medskip

\begin{equation}\label{Eq:1}
d = \frac{\sum_{t = \max{(d_{max}-t_l,1})}^{\min{(d_{max}+t_r,16})}t \hspace{0.1pt} \max{(0,h_t-b)}}{\sum_{t = \max{(d_{max}-t_l,1})}^{\min{(d_{max}+t_r,16})}\max{(0,h_t-b)}},
\end{equation}

\begin{figure}[h!] 
\centering\includegraphics[scale=0.33]{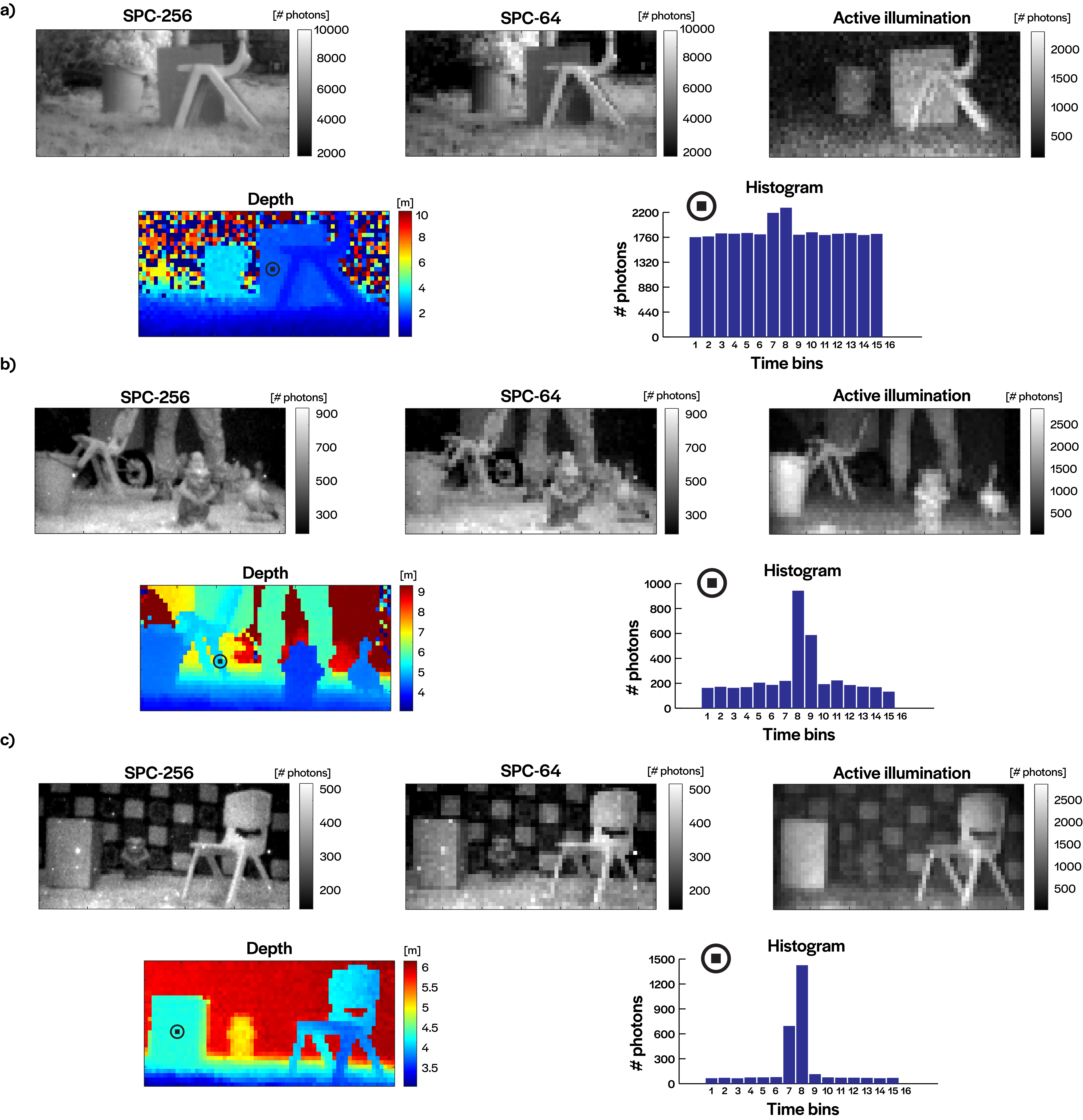}
\caption{SPC-256, SPC-64, Act\textunderscore I-D (active illumination representation only), depth and histogram examples for a) SBR < 0.1. (b) 0.1 < SBR < 0.5 and (c) SBR > 0.5. The histograms correspond to the indicated pixels on the depth maps (the pixel is shown in black and encircled). The last bin of the histogram (16) is not usable and is assigned a value of 0.}
\label{fig:Modes}
\end{figure}

\medskip

\noindent where $h_t$ is the histogram bin at a given macropixel (from 1 to 16), $d_{max}$ is the index of the bin with the maximum count, $b$ is the median of the bins (measure of ambient level) and $t_l$, $t_r$ are parameters with a value corresponding to the width of the histogram peak, typically $t_l$, $t_r$ = 2. In this approach, depth is estimated by calculating a centroid around the maximum count and its neighbours, providing sharp edges between objects in depth maps. Computing the centroid over all bins instead (after compensating for ambient level) would be akin to the depth map from an iToF camera, with the pixels seeing multiple returns outputting an averaged depth (so that edges between two objects appear blurrier). Data acquisition is carried out using an Opal Kelly XEM7310 FPGA integration module, which relays the SPAD data to a Matlab-based software interface, where histogram to depth conversion takes place. As in \cite{SPAD2}, we operate the camera in a hybrid modality, such that it alternates between histogram and intensity frames in a time-interleaved manner. In this paper, we consider a target detection neural network and investigate different data inputs based on the acquired data. Fig. \ref{fig:Modes} shows examples of the considered data types:

\begin{itemize}
    \item Intensity SPC-256 (256$\times$128), obtained under a combination of ambient and active illumination.
    \item Intensity SPC-64 (64$\times$32), resized from SPC-256.
    \item Active intensity and depth concatenated (Act\textunderscore I-D, 64$\times$32$\times$2). Active intensity is obtained by summing all the bins of the histogram (for a given pixel) after background subtraction.
    \item Depth (64$\times$32), obtained from histogram via Eq. \ref{Eq:1}.
    \item Histogram (64$\times$32$\times$16).
    
\end{itemize}

The three images given are representative of three different categories of signal-to-background ratio that we define here: very low (SBR < 0.1, Fig. 1a), low (0.1 < SBR < 0.5, Fig. 1b) and moderate (SBR > 0.5, Fig. 1c). The lower the SBR, the more difficult it is to extract the signal peak in the histogram, and eventually it gets buried within the ambient photon counts, leading to the noisy background seen in the depth image of Fig 1a.

\medskip

\medskip

We note that SPAD dToF sensors are commonly based on a "first photon" time-to-digital converter (TDC) architecture, whereby each pixel only registers the first detected photon per frame. This then leads to significant pile-up distortion (or exponential decay in the observed histogram) in high ambient conditions, leading to inaccurate depth maps \cite{Gupta_2019_ICCV,Rapp:21}. However, the SPAD used here instead features a multi-event TDC and is therefore more robust to strong ambient light, as detailed in \cite{SPAD2}.

\section{Object Detection via Neural Networks}

Amongst a number of methods for object detection, the use of deep learning has gained popularity in recent years. CNNs are powerful tools that are able to recognise key features of every object present in an image and differentiate them with a high accuracy. One way to perform this task is via segmentation, where every pixel in the image is assigned a certain class. U-net is a popular network that has proved to be effective in segmentation problems by using small datasets of hundreds of examples [26]. 

\medskip

\begin{figure}[h!] 
\centering\includegraphics[scale=0.155]{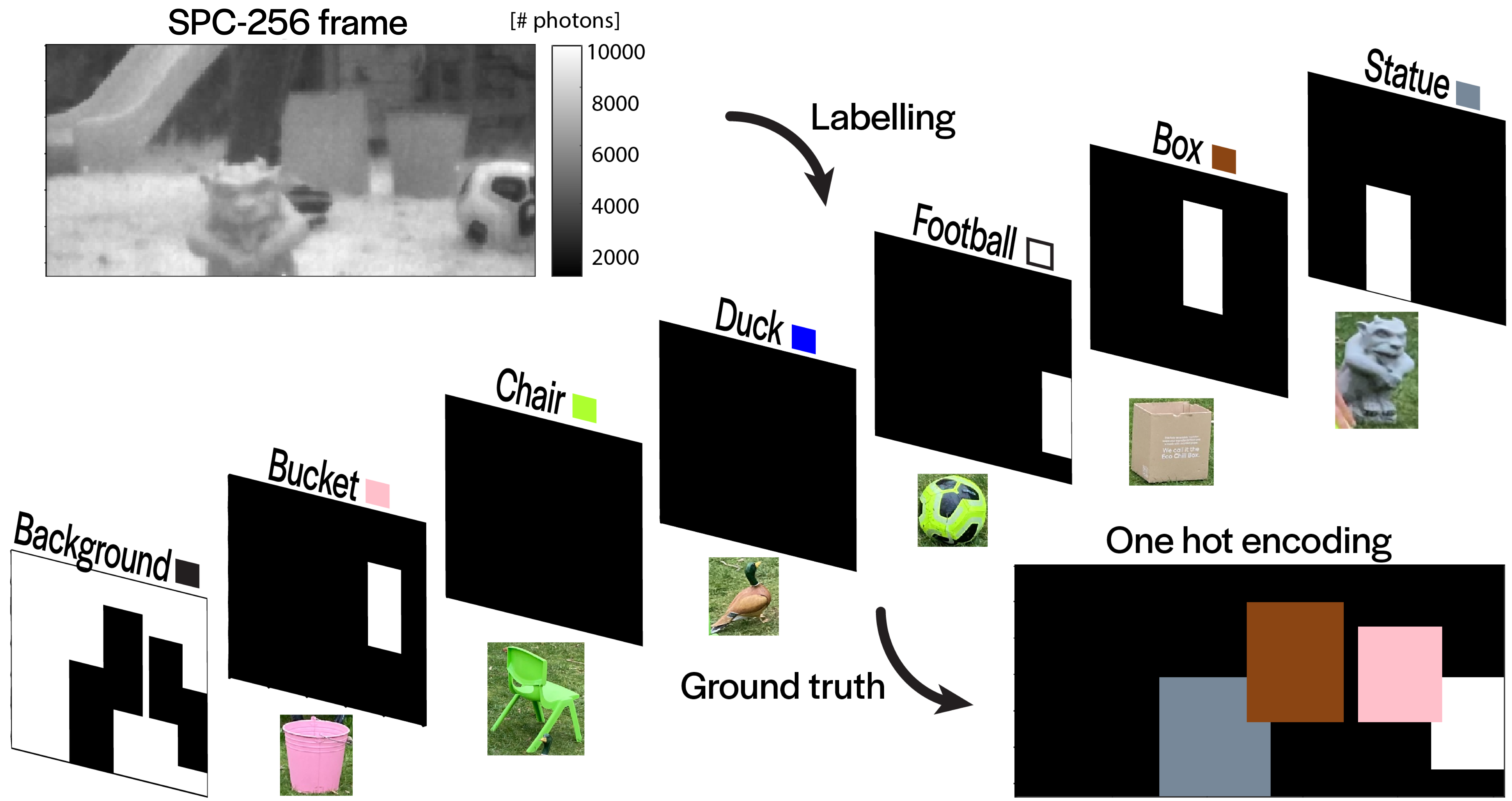}
\caption{Example of a SPC-256 frame, its corresponding ground truth (split for each class) and its one hot encoding (associating a color for each class). The ground truth has pixel values of 0 (black, no object) and 1 (white, object is present).}
\label{fig:NN_OHE}
\end{figure}

In this paper, a neural network with the main structure of U-net is used \cite{U-net}. Certain changes are applied with regards the size of input/output layers to match the size of our data: depth and SPC-64 (64$\times$32), Act\textunderscore I-D (64$\times$32$\times$2), SPC-256 (256$\times$128) and histogram (64$\times$32$\times$16). Table S5 on the Supplemental Document gives a full description of each layer of the neural network. The network is first trained on data with known ground truth, and then deployed to perform target detection on new data. During training, we provide the network with example frames together with their corresponding ground truth, here based on manually labelled SPC-256 images and resized for other data types. The ground truth has the same lateral resolution as the data type to be trained and X channels, where X is the number of objects to be detected plus the background. In this study, we have selected 6 objects (bucket, chair, duck, football, cardboard box and statue). Therefore, ground truths are of the size 64$\times$32$\times$7 for depth, Act\textunderscore I-D, histogram and SPC-64 and 256$\times$128$\times$7 for SPC-256. Ground truths are created by labelling each frame with bounding boxes around the objects of interest. Each channel of the ground truth's array is then filled with 0 except for labelled regions, which are filled with 1. Ground truths are one hot encoded to easily visualize each class, giving each class a different RGB colour. Fig. \ref{fig:NN_OHE} shows an example of a frame and its corresponding ground truth and one hot encoding.

\medskip

\medskip

Each frame must be pre-processed before being fed to the neural network. All data types are normalized to values between 0 and 1. Depth data is corrected with a calibration frame to compensate for temporal skew in photon timing across the SPAD array. SPC-256 data is processed with a median filter of size 2$\times$2 to remove any outliers in the frame. SPC-64 is a resized version of SPC-256. Histogram data is altered by subtracting the median of each pixel's histogram to remove the background level. When summing all the bins of this histogram, we get a type of intensity image that preserves the active illumination from the laser but has minimal contribution from ambient light. This active intensity data, together with the depth frame, are concatenated to form the data type Act\textunderscore I-D. The dataset as a whole is augmented by applying a horizontal flip to each frame, thus doubling the quantity of frames. Finally, the dataset's order is shuffled randomly to prevent tuning the weights of the neural network specifically for a given class before seeing examples of others. Not doing so can lead to a local minimum in the optimisation problem far from the absolute minimum.

\medskip

The neural network is implemented in Tensorflow using Keras \cite{Keras}. The training is performed using the Adam optimizer \cite{Adam} on a desktop computer (HP EliteDesk 800 G5 TWR) with the assistance of a RTX2070 GPU to accelerate the process. The loss function selected to optimise this segmentation problem is the Focal Tversky loss, often used when there is a class imbalance (typically the background being much more represented than other classes) \cite{Tversky}. The Focal Tversky loss or FTL follows the expression of Eq. (\ref{Eq:2})

\begin{equation} \label{Eq:2}
    FTL = \left( 1-\frac{TP}{TP+\alpha FN +\beta FP}  \right ) ^\gamma,
\end{equation}

\medskip

\noindent where TP is true positives, FN false negatives, FP false positives, $\alpha$ and $\beta$ are two parameters which $\alpha + \beta = 1$ and penalises false negatives when $\alpha > \beta$ and $\gamma$ is a parameter useful for class imbalance that forces the model to focus on harder examples when $\gamma > 1$. In this network, $\alpha = 0.6$, $\beta = 0.4$ and $\gamma = 1.2$. The metric to track the performance of the model is the F-score (F$_1$), calculated via Eq. (\ref{Eq:3}):

\begin{equation} \label{Eq:3}
   F_1 = \frac{TP}{TP+\frac{1}{2}(FP+FN)}.
\end{equation}

\medskip

Additional parameters to be considered for the neural network are the number of epochs and the batch size, set to 100 and 32, respectively. Early stopping of the training process is set with a patience of 8 using the minimum of the validation loss as metric. In other words, if the validation loss has not decreased after 8 epochs, the process is terminated and saves a model with the weights of the best training step. Once the training is complete, an evaluation of the performance of the model is required. In an object detection problem, we are interested in both localising and classifying correctly each class in the scene. An object is considered to be well localised when its intersection over union (IoU) surpasses 50\%. IoU is defined as the area of overlap between the predicted segmentation and the ground truth divided by the area of union between the predicted segmentation and the ground truth \cite{IoU}. Similarly, an object is considered to be correctly classified when the classes of the prediction and the ground truth are matching. The object is successfully detected only if both criteria are satisfied at the same time.

\section{Experimental Set-up}

\begin{figure}[h!] 
\centering\includegraphics[scale=0.48]{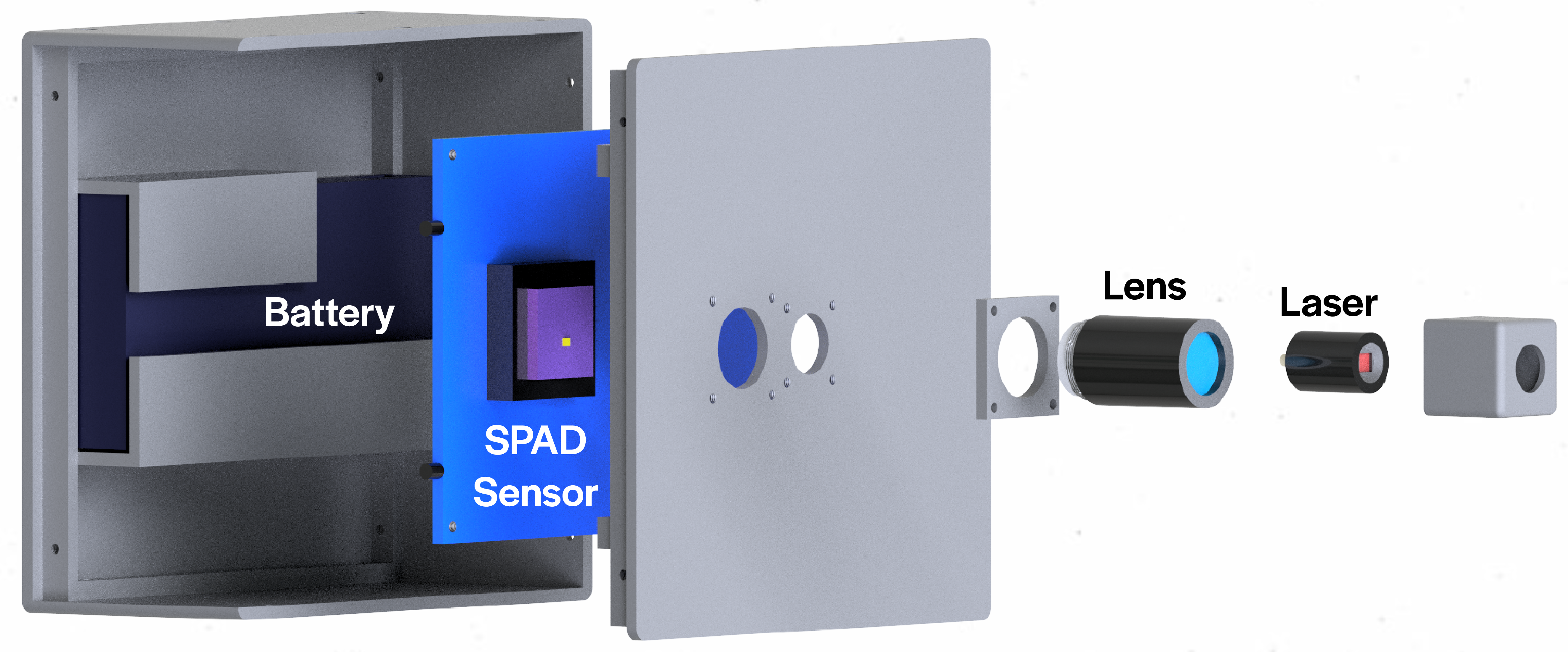}
\caption{Rendered image of the 3D-printed SPAD camera box. The camera contains the SPAD sensor, a lens, a 850 nm NIR laser and a battery for portability.}
\label{fig:SPAD_sensor}
\end{figure}

To facilitate the use of the SPAD outdoors, a portable camera setup is built, housed within a 180$\times$180$\times$88 mm 3D printed enclosure. Fig. \ref{fig:SPAD_sensor} shows a rendered image of the SPAD camera box with all the components.

\medskip

\medskip

The camera includes a NIR laser source (850 nm) that is triggered from the SPAD printed circuit board (PCB). This laser comes in a compact format (19 mm diameter and 28 mm long, Iberoptics Time-of-flight illuminator) and provides a peak power of 2 W distributed uniformly over a FoV of 20º. The laser emits 10 ns pulses with a repetition rate of 6 MHz, which makes it suitable for short range applications (up to 10 m). Next to the laser, a 6 mm focal length C-mount lens (MVL6WA) is placed to ensure matching illumination and imaging FoVs. The lens includes an ambient filter (Thorlabs FL850-10) to reduce the number of ambient photons that reach to the sensor. The box provides a hole compatible with tripod mounts and a portable battery that powers the camera. The SPAD sensor PCB is connected to a laptop via a USB3.0 connection to display the output in real-time, as explained in Section 2.

\section{Results}

We evaluate the performance of object detection for the data types offered by the SPAD camera: depth, SPC-256, SPC-64, Act\textunderscore I-D and histogram. To create the necessary dataset, a large number of frames is captured featuring the objects described in Section 3, which are placed at different positions and distances from the camera. Exposure times of 2ms, 5ms and 10ms are used. This is repeated for different ambient conditions and locations within a garden setting.

\medskip

\begin{figure}[ht!]
\centering\includegraphics[scale=0.38]{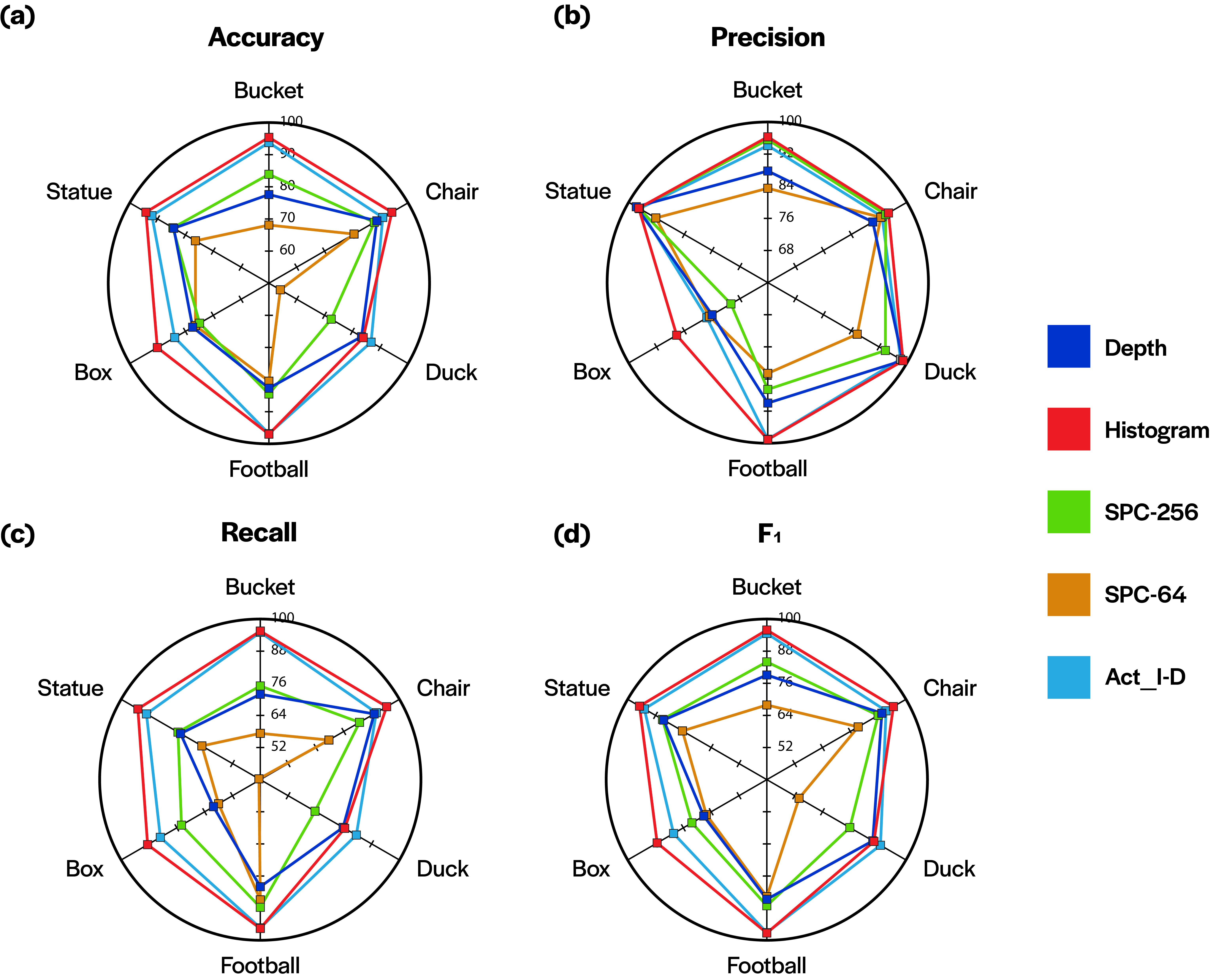}
\caption{Statistics for depth, histogram, SPC-256, SPC-64 and Act\textunderscore I-D: (a) Accuracy. (b) Precision. (c) Recall. (d) F-score.}
\label{fig:Spider_6_9}
\end{figure}

In the first trial, a total of 6388 training examples are fed into the neural network to create a model and a total of 574 examples are used for testing with a SBR < 0.5 on average (the test data is drawn from a separate subset of the dataset, with a different backdrop). A validation set, corresponding to 15\% of the data from the training set is used to provide and unbiased evaluation of the model. A RTX2070 GPU is used to speed-up the training and prediction tasks involved in the neural network. This implementation boosts processing speeds up to 1325 FPS for histograms, 1100 FPS for Act\textunderscore I-D, 1780 FPS for both depth and SPC-64 and 280 FPS for SPC-256, which makes it compatible with high-speed vision. The training is run 10 times since the optimization of the network's weights differs every time, leading to changes in the predictions; multiple runs thus enable statistical analyses to be performed. Four parameters are used to assess the network's performance in this study. The accuracy reflects the ratio between correct predictions and total observations. Precision measures the ratio of correct positive predictions and total predicted positive observations. Recall, or also known as sensitivity, measures the ratio of correct positive predictions and all observations in a class. Finally, the F-score (Eq. \ref{Eq:3}) is also measured since it is a weighted average of precision and recall. This parameter becomes meaningful specially when the amount of false positives and false negatives is not balanced. Fig. \ref{fig:Spider_6_9} summarizes the average of the parameters described above for each data type and object. A table with all values can be found on the Supplemental Document (Table S1).

\medskip

As seen in Fig. \ref{fig:Spider_6_9}, the data suggests that even when capturing frames under high ambient conditions, the overall object detection accuracy consistently surpasses 70\% for all objects and data types (>80\% for histogram and Act\textunderscore I-D), except for the case of SPC-64. The latter data type has low lateral resolution, so the contrast between the objects and what is often a complex background can be poor, making it very challenging to locate and classify objects accurately. In terms of the performance differences between objects, these are thought to arise mainly due to their relative shapes and sizes. The football has one of the highest performance in most cases thanks to its simple round shape, whereas the duck presents lower figures given its more complex shape which varies significantly with orientation.

\medskip

High precision values indicate that the neural network is able to identify the right class wherever there is an object, or in other words, there is a low incidence of false positives. SPC-64 presents again the lowest numbers for most of the objects and histogram gives the most consistently high precision. Recall values are in general lower than precision ones, indicating a higher level of false negatives in the examples. False negatives refer to cases which either there are detections which do not appear in the ground truth or if they do, the IoU is below 50\%. Given the significant difference between the incidence of false positives and false negatives, the F-score represents a more balanced assessment than accuracy when comparing the relative performance of the different data types for the various objects. Overall, histograms have the best score (92.7\% average F-score) closely followed by Act\textunderscore I-D (90.9 \% average F-score). As before, SPC-64 has the worst performance (71.3\% average F-score)

\medskip

The F-score values are submitted to a t-student test to assess if the performance of different data types is significantly different from each other (considering the variability in performance every time the network is trained). Table \ref{tab:test_6_9} summarises the results of the tests looking for statistically significant differences in performance between histograms and the other data types after 20 runs. Histograms are seen to have a clear advantage over other data types, apart from Act\textunderscore I-D. Comparisons between other data types can be found on Table S2 in the Supplemental Document.

\begin{table}[htbp]
\centering
\caption{\bf T-student analysis of F-score results of histogram versus other data types. Green - histogram has higher F-score, yellow - no difference, red - other data type has higher F-score.}
\resizebox{100mm}{!}{%

\begin{tabular}{cllll}

\hline
 & \multicolumn{1}{c}{\textbf{Hist > Depth}} & \multicolumn{1}{c}{\textbf{Hist > SPC-256}} & \multicolumn{1}{c}{\textbf{Hist > SPC-64}} & \multicolumn{1}{c}{\textbf{Hist > Act\textunderscore I-D}} \\ \hline
\textbf{Bucket} & \cellcolor[HTML]{C6EFCE}{\color[HTML]{006100} } & \cellcolor[HTML]{C6EFCE}{\color[HTML]{006100} } & \cellcolor[HTML]{C6EFCE}{\color[HTML]{006100} } & \cellcolor[HTML]{FFEB9C}{\color[HTML]{9C6500} } \\
\textbf{Chair} & \cellcolor[HTML]{C6EFCE}{\color[HTML]{006100} } & \cellcolor[HTML]{C6EFCE}{\color[HTML]{006100} } & \cellcolor[HTML]{C6EFCE}{\color[HTML]{006100} } & \cellcolor[HTML]{C6EFCE}{\color[HTML]{006100} } \\
\textbf{Duck} &  \cellcolor[HTML]{FFEB9C}{\color[HTML]{9C6500} } & \cellcolor[HTML]{C6EFCE}{\color[HTML]{006100} } & \cellcolor[HTML]{C6EFCE}{\color[HTML]{006100} } & \cellcolor[HTML]{FFC7CE}{\color[HTML]{9C0006} } \\
\textbf{Football} & \cellcolor[HTML]{C6EFCE}{\color[HTML]{006100} } & \cellcolor[HTML]{C6EFCE}{\color[HTML]{006100} } & \cellcolor[HTML]{C6EFCE}{\color[HTML]{006100} } & \cellcolor[HTML]{FFEB9C}{\color[HTML]{9C6500} }  \\
\textbf{Box} & \cellcolor[HTML]{C6EFCE}{\color[HTML]{006100} } & \cellcolor[HTML]{C6EFCE}{\color[HTML]{006100} } & \cellcolor[HTML]{C6EFCE}{\color[HTML]{006100} } & \cellcolor[HTML]{C6EFCE}{\color[HTML]{006100} } \\
\textbf{Statue} & \cellcolor[HTML]{C6EFCE}{\color[HTML]{006100} } & \cellcolor[HTML]{C6EFCE}{\color[HTML]{006100} } & \cellcolor[HTML]{C6EFCE}{\color[HTML]{006100} } & \cellcolor[HTML]{C6EFCE}{\color[HTML]{006100} } \\ \hline
\end{tabular}%
}
  \label{tab:test_6_9}
\end{table}

\medskip

It is interesting to observe histograms outperforming SPC-256, despite the higher lateral resolution of the latter (and the high ambient levels benefiting the intensity data, whilst adding noise to histograms). This indicates that time resolved imaging can capture salient details in a scene even when the x,y resolution is coarse. In practice, the performance of intensity-based object detection is expected to depend on the level of contrast in colour and texture between the objects and the backdrop. Thus, perhaps a more significant observation is histogram-based (and Act\textunderscore I-D) processing being considerably more effective than using depth alone. This points to active intensity data (equivalent to the number of photons in the histogram peak) providing a rich set of information. This represents a fine projection of the information available in histogram data, which can be used for faster processing of results with little compromise on the detection performance. Cases where histograms outperform Act\textunderscore I-D may be explained by the multi-peak content for pixels at the edges of objects, which may help in the perception of the contour of objects with higher fidelity.

\medskip

A closer examination of detections is carried out with the same test dataset to try and establish patterns in the failure cases. Histogram data (which performs the best) is mutually compared with other data types (e.g histogram vs depth), by analysing the percentage of examples where both predict correctly, where both are wrong and where only one predicts correctly. 

\begin{table}[htbp]
\centering
\caption{\bf Failed detection comparison between histograms and other data types. The parameters are: \% of examples where both histogram and the other data type predict correctly, where both are incorrect, and where one is correct and the other incorrect.}

\resizebox{\textwidth}{!}{%
\begin{tabular}{ccccc|cccc|cccc|cccc}
\hline
 & \multicolumn{4}{c}{\textbf{Hist vs Depth}} & \multicolumn{4}{c}{\textbf{Hist vs SPC-256}} & \multicolumn{4}{c}{\textbf{Hist vs SPC-64}} & \multicolumn{4}{c}{\textbf{Hist vs Act\textunderscore I-D}} \\
 & \textbf{Both \cmark} & \textbf{Both \xmark} & \textbf{Hist \cmark} & \multicolumn{1}{c}{\textbf{Depth \cmark}} & \textbf{Both \cmark} & \textbf{Both \xmark} & \textbf{Hist \cmark} & \multicolumn{1}{c}{\textbf{SPC-256 \cmark}} & \textbf{Both \cmark} & \textbf{Both \xmark} & \textbf{Hist \cmark} & \multicolumn{1}{c}{\textbf{SPC-64 \cmark}} & \textbf{Both \cmark} & \textbf{Both \xmark} & \textbf{Hist \cmark} & \multicolumn{1}{c}{\textbf{Act\textunderscore I-D \cmark}} \\ \hline
\textbf{Bucket} & 90.6 & 0.0 & 8.5 & 0.9 & 96.9 & 0.0 & 2.2 & 0.9 & 77.5 & 0.0 & 21.6 & 0.9 & 98.4 & 0.0 & 0.6 & 0.9 \\
\textbf{Chair} & 93.1 & 0.3 & 5.3 & 1.3 & 95.7 & 0.3 & 2.6 & 1.3 & 89.1 & 0.7 & 9.2 & 1.0 & 91.4 & 0.0 & 6.9 & 1.6 \\
\textbf{Duck} & 93.9 & 0.3 & 3.4 & 2.4 & 82.5 & 0.3 & 14.8 & 2.4 & 49.5 & 1.1 & 47.9 & 1.6 & 93.9 & 0.8 & 3.4 & 1.9 \\
\textbf{Football} & 91.6 & 0.0 & 6.8 & 1.6 & 93.8 & 0.0 & 4.7 & 1.6 & 81.7 & 0.0 & 16.8 & 1.6 & 97.5 & 0.0 & 0.9 & 1.6 \\
\textbf{Box} & 89.6 & 0.0 & 5.2 & 5.2 & 85.8 & 0.5 & 9.0 & 4.7 & 81.6 & 0.0 & 13.2 & 5.2 & 89.6 & 0.5 & 5.2 & 4.7 \\
\textbf{Statue} & 96.4 & 0.6 & 1.8 & 1.2 & 93.3 & 0.0 & 4.8 & 1.8 & 93.0 & 0.3 & 5.2 & 1.5 & 96.7 & 0.6 & 1.5 & 1.2 \\ \hline
\end{tabular}
}
\label{tab:misclassifications}
\end{table}

\begin{figure}[ht!]
\centering\includegraphics[scale=0.2]{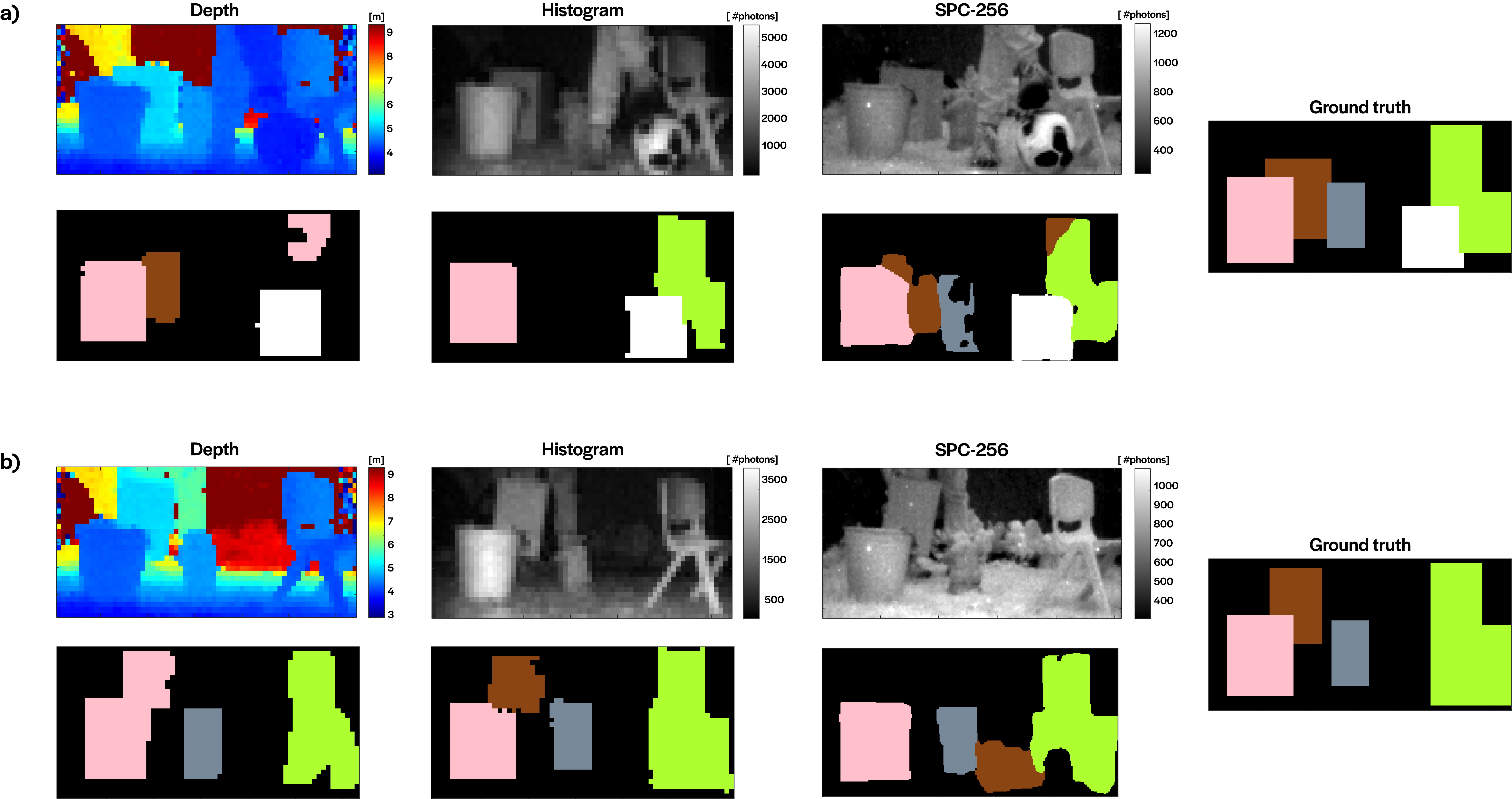}
\caption{Misclassification examples (a) and (b). Depth, histogram (active illumination representation) and SPC-256 frames with its corresponding one hot encoding prediction below. On the rightmost, the ground truth of the given example. }
\label{fig:Misclassifications}
\end{figure}

On Table \ref{tab:misclassifications} we see that the vast majority of examples are predicted correctly by either one or both of the paired data types. It is only in a minority of cases (typically <1\%) that both data types fail to detect and object at the same time. The results suggest that the different data types provide in some ways complementary information on the scene, and there may be potential benefits in combining data types for object detection. Fig.  \ref{fig:Misclassifications} shows challenging examples where the neural network fails to detect appropriately from depth, histogram (represented as active illumination) and SPC-256. In the first example, all data types have at least one false detection. In the second example, the histogram prediction detects all objects correctly, whereas depth and SPC-256 have some failed detections. It can be seen how the examples have a complex background and contain multiple objects (some of them partially occluded) which leads to a lower prediction performance.

\medskip

\medskip

It is also of interest to assess the impact of SBR being further reduced on the neural network's performance. For this study, the dataset is partitioned in a different way to enable a subset with lower SBR to be used for testing. The model is trained with 4086 examples and tested with 1151 examples, now with a SBR < 0.05 on average (i.e. 10$\times$ lower SBR than before). The discussion of these results can be found in the Supplemental Document, containing a table and figure with all values (Table S3 and Figure S1). We note that due to the considerably lower SBR than before, histograms are significant noisier, making histogram-based (and by extension depth-based) perception more challenging. Nevertheless, relative performance of the different data types is broadly similar, with histograms still offering the highest, and most consistent performance overall (85.2 \% F-score on average). Act\textunderscore I-D again produces similar results to histograms (82.7 \% F-score on average). 

\medskip

We also consider object detection of a high-speed moving object under strong ambient conditions (SBR < 0.05), in this case a football. Several sequences totaling 245 frames are taken at 500 FPS (2 ms exposure time) and tested with the models generated by the complete dataset (6388 examples, none of them including high-speed sequences). The video of the sequence can be found in the supplemental material (Visualization 1). Table \ref{tab:HS_football} shows the average parameters for football detection for all data types.

\begin{table}[htbp]
\centering
\caption{\bf Average performance parameters after 10 neural network training runs for a football moving at high-speed as captured by different data types: depth, histogram, SPC-256, SPC-64 and Act\textunderscore I-D.}

\begin{tabular}{ccccc}
\hline
 & \textbf{Accuracy  [\%]} & \textbf{Precision [\%]} & \textbf{Recall [\%]} & \textbf{F$_1$ [\%]} \\ \hline
\textbf{Depth} & 94.3 & 100 & 78.8 & 88.1 \\
\textbf{Hist} & 98.0 & 100 & 92.4 & 96.1 \\
\textbf{SPC-256} & 99.6 & 100 & 98.5 & 99.2 \\
\textbf{SPC-64} & 98.8 & 100 & 95.5 & 97.7 \\
\textbf{Act\textunderscore I-D} & 97.1 & 100 & 89.4 & 94.4 \\ \hline
\end{tabular}%

  \label{tab:HS_football}
\end{table}

\bigskip
Fast moving objects may contain motion blur when they are captured (even at high frame rates), meaning that the object presented to the neural network might look unfamiliar and remain undetected. This issue combined with low lateral resolution can complicate the detection performance. However, the detection of a high-speed moving football appears to be successful for all data types, with histogram-based processing again outperforming depth.

\section{Conclusion}

This paper presents the application of a dToF SPAD sensor to high-speed object detection in outdoor conditions. A dataset involving six different objects located at various positions, with different backdrops, and under varying ambient levels is created to train and test neural networks for different data types. These include depth, histogram, Act\textunderscore I-D (intensity under active illumination, combined with depth), and two types of intensity; SPC-256 and SPC-64 (resized from SPC-256). Processing times are commensurate with the high-speed operation of the camera, which runs at hundreds of frames per second.

\medskip

The results from test datasets with signal to background ratios (SBR) of <0.5 and <0.05 indicate that object detection based on histogram data gives the highest performance overall, with average F-scores of 92.7\% and 85.2\%, respectively. Act\textunderscore I-D offers a similar performance, with F-score averages of 90.9\% and 82.7\%, respectively. This compares with 81.8\% and 69.7\% for inference based on depth alone.

\medskip

We note that since the advantage of histogram data versus Act\textunderscore I-D appears to be relatively modest, for dToF SPADs with larger number of bins than the 16 bins here, Act\textunderscore I-D may be preferable to avoid the need for large neural network models. 

\medskip

We believe that the results presented here are relevant for high-speed situational awareness in autonomous systems. Whilst these systems often use the fusion of an assortment of sensors to interpret the surroundings, for the limited number of objects (6), and short range (<10m) considered here, the SPAD on its own shows a promising level of performance. It would be interesting to expand the study to more generalised conditions, and to consider, for example, the addition of RGB data.

\section*{Funding}

This work was supported by EPSRC through grants EP/M01326X/1 and EP/S001638/1. Also it is supported by the UK Royal Academy of Engineering Research Fellowship Scheme (Project RF/201718/17128) and DSTL Dasa project (DSTLX1000147844).

\section*{Acknowledgements}

The authors are grateful to STMicroelectronics and the ENIAC-POLIS project for chip fabrication. Preliminary results from this work, using a different neural network model with slower processing, and indoor data only, were presented at Photonic Instrumentation Engineering VIII in 2021 \cite{martin2021high}.

\section*{Disclosures}

The authors declare no conflicts of interest.

\section*{Supplemental Document}

See Supplement 1 for supporting content.


\bibliography{references}

\end{document}